\documentclass[prd,twocolumn,a4paper,superscriptaddress,floatfix]{revtex4}
\usepackage{graphicx}
\usepackage{bbm}
\usepackage[all]{xy}
\usepackage{amsmath}
\usepackage{amssymb}
\usepackage{epstopdf}

\def\be{\begin{equation}}
\def\ee{\end{equation}}
\newcommand{\bq}{\begin{eqnarray}}
\newcommand{\eq}{\end{eqnarray}}
\newcommand{\bes}{\begin{subequations}}
\newcommand{\ees}{\end{subequations}}

\def\ben{\begin{eqnarray}}
\def\een{\end{eqnarray}}
\def\ba{\begin{array}}
\def\ea{\end{array}}

\begin{document}
\newcommand{\half}{{\textstyle\frac{1}{2}}}
\allowdisplaybreaks[3]
\def\a{\alpha}
\def\b{\beta}
\def\g{\gamma}\def\G{\Gamma}
\def\d{\delta}\def\D{\Delta}
\def\ep{\epsilon}
\def\et{\eta}
\def\z{\zeta}
\def\t{\theta}\def\T{\Theta}
\def\l{\lambda}\def\L{\Lambda}
\def\m{\mu}
\def\f{\phi}\def\F{\Phi}
\def\n{\nu}
\def\p{\psi}\def\P{\Psi}
\def\r{\rho}
\def\s{\sigma}\def\S{\Sigma}
\def\ta{\tau}
\def\x{\chi}
\def\o{\omega}\def\O{\Omega}
\def\k{\kappa}
\def\pa {\partial}
\def\ov{\over}
\def\br{\\}
\def\ud{\underline}

\newcommand\lsim{\mathrel{\rlap{\lower4pt\hbox{\hskip1pt$\sim$}}
    \raise1pt\hbox{$<$}}}
\newcommand\gsim{\mathrel{\rlap{\lower4pt\hbox{\hskip1pt$\sim$}}
    \raise1pt\hbox{$>$}}}
\newcommand\esim{\mathrel{\rlap{\raise2pt\hbox{\hskip0pt$\sim$}}
    \lower1pt\hbox{$-$}}}

\title{Impact of string and monopole-type junctions on domain wall dynamics: implications for dark energy}

\author{L. Sousa}
\email[Electronic address: ]{laragsousa@gmail.com}
\affiliation{Centro de F\'{\i}sica do Porto, Rua do Campo Alegre 687, 4169-007 Porto, Portugal}
\affiliation{Departamento de F\'{\i}sica da Faculdade de Ci\^encias
da Universidade do Porto, Rua do Campo Alegre 687, 4169-007 Porto, Portugal}
\author{P.P. Avelino}
\email[Electronic address: ]{ppavelin@fc.up.pt}
\affiliation{Centro de F\'{\i}sica do Porto, Rua do Campo Alegre 687, 4169-007 Porto, Portugal}
\affiliation{Departamento de F\'{\i}sica da Faculdade de Ci\^encias
da Universidade do Porto, Rua do Campo Alegre 687, 4169-007 Porto, Portugal}

\begin{abstract}

We investigate the potential role of string and monopole-type junctions in the frustration of domain wall networks using a velocity-dependent one-scale model for the characteristic velocity, $v$, and the characteristic length, $L$, of the network. We show that, except for very special network configurations, $v^2 \lsim (HL)^2 \lsim (\rho_\sigma + \rho_\mu)/\rho_m$ where $H$ is the Hubble parameter and $\rho_\sigma$, $\rho_\mu$ and $\rho_m$ are the average density of domain walls, strings and monopole-type junctions. We further show that if domain walls are to provide a significant contribution to the dark energy without generating exceedingly large CMB temperature fluctuations then, at the present time, the network must have a characteristic length $ L_0 \lsim 10  \,\Omega_{\sigma 0}^{-2/3} \, {\rm kpc}$ and a characteristic velocity $v_0 \lsim 10^{-5} \, \Omega_{\sigma 0}^{-2/3}$ where $\Omega_{\sigma 0}=\rho_{\sigma 0}/\rho_{c 0}$ and $\rho_c$ is the critical density. In order to satisfy these constraints with $\Omega_{\sigma 0} \sim 1$, $\rho_{m 0}$ would have to be at least $10$ orders of magnitude larger than $\rho_{\sigma 0}$, which would be in complete disagreement with observations. This result provides very strong additional support for the conjecture that no natural frustration mechanism, which could lead to a significant contribution of domain walls to the dark energy budget, exists.

\end{abstract} 
\maketitle

\section{Introduction}

There is now very strong observational evidence that our Universe is currently undergoing a period of accelerated expansion \cite{Komatsu:2008hk,Frieman:2008sn}. In the framework of General Relativity, the observed acceleration of the Universe ought to be explained by the existence of an exotic dark energy component which violates the strong energy condition. In  \cite{Bucher:1998mh} it was suggested, for the first time, that a frozen domain wall network could be responsible for such acceleration and 
several other authors have subsequently advocated this possibility \cite{Carter:2004dk,Battye:2005hw,Battye:2005ik,Carter:2006cf}.
 
The evolution of cosmological domain wall networks has been studied in detail using both high-resolution numerical simulations and a semi-analitical velocity-dependent one-scale (VOS) model \cite{PinaAvelino:2006ia,Avelino:2006xy,Avelino:2006xf,Battye:2006pf,Avelino:2008ve,Avelino:2009tk}. Strong analytical and numerical evidence was provided that a frustrated domain wall network, accounting for a significant fraction of the energy density of the Universe at the present time, will never emerge from realistic phase transitions. These results appear to rule out a significant contribution of domain walls to the dark energy budget.

In all these studies the energy of the domain walls was assumed to be very small. Still, domain wall junctions can be responsible for freezing a domain wall network if they are heavy enough. In that case their contribution to the energy budget could not be neglected but it has been argued in \cite{Avelino:2006xf,Avelino:2008ve} that it would spoil the dark energy properties associated with a frustrated domain wall network. More recently, a mechanism for dynamical frustration of domain wall networks involving kinky vortons was suggested in  \cite{Battye:2009ac}.

In this paper we investigate the impact of string and monopole-type junctions on the dynamics of domain wall networks. Accounting for the detailed contribution of the junctions on domain wall network simulations is not a trivial task. In fact, the standard Press-Ryden-Spergel (PRS) algorithm \cite{Press} (implemented in all field theory numerical simulations in order to ensure fixed comoving resolution) increases artificially the impact of the junctions on the overall network dynamics during the course of the simulations. This effect is not very important for the light junctions which are usually considered in such simulations but could be relevant in the case of heavy junctions. 

In order to overcome this potential problem we develop a semi-analytical VOS model which incorporates  the contribution of string and monopole-type junctions to the overall dynamics of domain wall networks. We use this model to constrain the characteristic length and velocity of domain wall networks and discuss the corresponding cosmological implications, in particular for dark energy.

\section{Semi-analitical VOS model with junctions}

\begin{figure}[t!]
\includegraphics[width=3.1in]{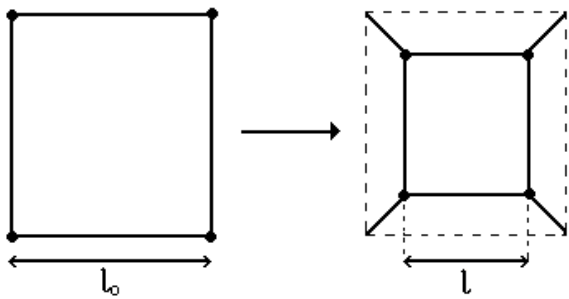}
\caption{\label{junctions} The left panel represents the domain wall configuration prior to collapse. Solid lines represent domain walls of superficial tension $\sigma$ and the large dots represent string junctions of tension $\mu$. The right panel represents the configuration after some time has elapsed.}
\end{figure}

Consider the configuration presented in figure \ref{junctions} where 4 straight string-type junctions of energy per unit length $\mu$ (represented by 4 large dots) are connected by 4 planar domain walls of energy per unit length $\sigma$ (represented by 4 straight lines) defining a square domain of characteristic size $l$ ($l_0$ represents the characteristic size prior to collapse). Here we assume that nothing varies along the direction perpendicular to the square domain, so that the infinite string-type junctions have no curvature. In this case the domain wall dynamics is effectively 2-dimensional. The energy per unit length of this configuration is given by 
\be
E=4\sigma\gamma l+4\mu\gamma_\mu+2 {\sqrt 2} \sigma(l_0-l)\,,
\ee
where $\gamma=\gamma_\sigma=(1-v^2)^{-1/2}$, $\gamma_\mu=(1-v_\mu^2)^{-1/2}$, $v=v_\sigma=-(dl/dt)/2$ represents the domain wall velocity and $v_\mu=\sqrt{2}v$ is the velocity of the string-type junctions. 

For simplicity, we shall make the approximation that the shape is maintained during the collapse of the square domain so that the 
curvature of the domain walls remains concentrated at the vertices. Although we expect domain walls to acquire curvature in the course of the collapse, this approximation will capture the relevant physics and does not affect our main results. It follows from energy conservation that, in Minkowski spacetime, the equation of motion is given by
\be
\frac{dv}{dt}-(1-v^2)\frac{s(v)}{l}\left(\frac{1}{1+2\frac{\mu}{\sigma l}g(v)}\right)=0\,,
\label{micro}
\ee
where 
\be
s(v)=2-\frac{\sqrt 2}{\gamma}\,,
\ee
and
\be
g(v)=\left(\frac{\gamma_\mu}{\gamma}\right)^3\,.
\ee
These results do not directly apply to the case of a domain wall network since they were obtained for a very particular configuration. Still, we shall use Eq. (\ref{micro}) as a guide into the fundamental aspects of network evolution.

The characteristic lengths, $L_\mu$ and $L_\sigma$, of a two-dimensional domain wall network with junctions are defined as
\be
\rho_\mu=\frac{\mu}{L_\mu^2}, \qquad \rho_{\sigma}=\frac{\sigma}{L_{\sigma}}\,,
\label{densities}
\ee
where $\rho_\mu$ and $\rho_\sigma$ represent the average density of string-type junctions and the average domain wall density, respectively. The relation between $L_\mu$ and $L_\sigma$ depends on the geometrical properties of the domain wall network. For a regular hexagonal lattice with Y-type junctions $L_\mu/L_\sigma=3^{1/4}$, for a regular square lattice with X-type junctions 
$L_\mu/L_\sigma=2$, while $L_\mu/L_\sigma=2^{1/2} \, 3^{3/4}$ for a regular triangular lattice with $*$-type junctions. In general, $L_\mu/L_\sigma \sim 1$ and, consequently, in the following we shall assume that $L=L_\mu=L_\sigma$. The characteristic velocities are also expected to be similar and consequently we shall make the approximation that $v=v_\mu=v_\sigma$.

Then substituting $l=L/2$ in Eq.  (\ref{micro}) one obtains, in the non-relativistic limit,
\be
\frac{dv}{dt}-\frac{k}{L} \left(\frac{1}{1+\frac{\mu}{\sigma L}}\right)=0\,,
\label{vos1}
\ee
with $k=2s(0)=1-{\sqrt 2}/2$. Note that the domain wall configuration presented in Fig. 1 is very specific. In general the curvature parameter, $k$,  depends on the geometrical properties of the domain wall network \cite{Avelino:2008ve}.

Eq (\ref{vos1}) can also be re-written as
\be
\frac{dv}{dt}-\frac{k}{L}(1-f_{\mu})=0\,,
\label{vos1'}
\ee
where $f_{\mu}=\rho_{\mu}/\rho$ and $\rho=\rho_{\sigma}+\rho_{\mu}$.

The generalization to the 3-dimensional case is trivial. In this case we need to consider the impact of monopole-type junctions, of mass $m$ and average density $\rho_m=m/L_m^3$, on the network dynamics. The evolution equation for the characteristic velocity becomes, up to coefficients of order unity,
\be
\frac{dv}{dt}-\frac{k}{L}\left(1-f_m\right)=0
\label{vos2'}
\ee
with $L=L_m=L_\mu=L_\sigma$ and $v=v_m=v_\mu=v_\sigma$. The energy density fractions of the various components  are
\bq
f_{\sigma}&=&\frac{\rho_{\sigma}}{\rho}=\frac{1}{1+\frac{m}{\sigma L^2} +\frac{\mu}{\sigma L}}\,,\\
f_{\mu}&=&\frac{\rho_{\mu}}{\rho}=\frac{1}{1+\frac{m}{\mu L} +\frac{\sigma L}{\mu}}\,,\\
f_m&=&\frac{\rho_{m}}{\rho}=\frac{1}{1+\frac{\mu L}{m} +\frac{\sigma L^2}{m}}\,,
\eq
where $\rho=\rho_\sigma+\rho_\mu+\rho_m$.

We have implicitly assumed that the energy of the domain walls as well as  that of string and monopole type junctions is localized, which may not always be a good approximation. For example, in the case of global monopoles described by a Lagrangian with a standard kinetic term there is a nearly constant long-range force between monopole and anti-monopole pairs which could be dynamically relevant. However, such long-range forces, if they are important, would trigger a very effective annihilation between monopole and anti-monopole pairs. This would constitute an additional obstacle to the frustration of a domain wall network and consequently we shall not consider them in the present paper. Long range forces between monopoles can be eliminated by considering non-standard kinetic terms which localize the energy of the monopoles inside their core.

In an expanding universe we  also need to account for the deceleration caused by the Hubble expansion. In the case of planar domain walls, the momentum per comoving area is proportional to $a^{-1}$, so that 
\be
\gamma_\sigma v_\sigma \propto a^{-3} \Leftrightarrow \frac{d v_\sigma}{dt}+3v_\sigma H(1-v_\sigma^2)=0\,.
\ee
Similarly, for cosmic strings the momentum per comoving length is proportional to $a^{-1}$. Consequently
\be
\gamma_\mu v_\mu \propto a^{-2} \Leftrightarrow \frac{d v_\mu}{dt}+2 v_\mu H(1- v_\mu^2)=0\,.
\ee
For massive junctions one has
\be
\gamma_m v_m \propto a^{-1} \Leftrightarrow \frac{d v_m}{dt}+v_m H(1-v_m^2)=0\,.
\ee

In the absence of energy loss mechanisms the evolution of the average energy density is given by
\bq
\rho_{\sigma}\propto\gamma_\sigma a^{-1}&\Leftrightarrow&\frac{d\rho_{\sigma}}{dt}+H(1+3v_\sigma^2)\rho_{\sigma}=0\,,\label{rho1}\\
\rho_{\mu}\propto\gamma_\mu a^{-2}&\Leftrightarrow&\frac{d\rho_{\mu}}{dt}+2(1+v_\mu^2)H\rho_{\mu}=0\,,\label{rho2}\\
\rho_{m}\propto\gamma_m a^{-3}&\Leftrightarrow&\frac{d\rho_{m}}{dt}+3\left(1+\frac{v_m^2}{3}\right)H\rho_{m}=0\,,
\label{rho3}
\eq
for domain walls, cosmic strings and point masses. In the course of network evolution energy will be exchanged between the different components, leading to similar characteristic velocities ($v_m \sim v_\mu \sim v_\sigma$).  To account for 
that one needs to include extra terms $Q_{\sigma}$, $Q_\mu$ and $Q_m$ on the right hand side of Eqs. (\ref{rho1}-\ref{rho3}), respectively. Here, $Q_X$ represents the energy transfered per unit of time and volume from the component $X$ to the other two components (note that $Q_m+Q_\mu+Q_\sigma=0$).

The evolution of the total energy density is given by
\be
\frac{d\rho}{dt}+\left[\left(1+f_{\mu}+2 f_{\sigma}\right)+v^2\left(3-f_{\mu}-2 f_\sigma\right)\right]H\rho=0\,,
\label{rhot}
\ee
where we have used  Eqs. (\ref{rho1}-\ref{rho3}), taking into account that $\rho=\rho_{\sigma}+\rho_{\mu}+\rho_m$ and the 
fact that the $Q_X$ terms cancel out.

Hence,
\be
\frac{dL}{dt}=\left[1+v^2\frac{1+f_{\mu}+2 f_{\sigma}}{3-f_{\mu}-2f_{\sigma}}\right]HL\,,
\label{rho4}
\ee
so that for $v=v_\sigma=v_\mu=v_m \ll 1$ one has $d L/dt = HL$ ($L \propto a$) with $L=L_\sigma=L_\mu=L_m$. A term proportional to $v \rho  /L$ could be added to the right hand side of Eq. (\ref{rhot}), in order to account for energy losses 
by the network. In the non-relativistic limit ($v \ll 1$) this term is negligible. In any case, it is easy to verify that this term leads to a larger $L$ and, consequently, it does not help frustration.

We use the energy density fractions as weight factors in the estimate of the Hubble damping term which needs to be added to Eq.  (\ref{vos1'}). The equation describing the evolution of the characteristic velocity of the network then becomes
\be
\frac{dv}{dt}+Hv(1+f_\mu+2 f_\sigma)-\frac{k}{L} \left(\frac{1}{1+\frac{m}{\mu L + \sigma L^2}}\right)=0\,,
\label{vos3}
\ee

Solving Eq.  (\ref{vos3}) in the matter dominated era, assuming that $\rho_\mu = 0$ and $\rho_m \gg \rho_\sigma$ and neglecting the decaying mode, one obtains
\be
H v = \frac{2}{5} \frac{d v}{dt} = \frac{2}{7} \frac{k}{L} \frac{\rho_\sigma}{\rho_m}\,.
\ee
A similar analysis for $\rho_\sigma = 0$ and $\rho_m \gg \rho_\mu$ would give
\be
H v = \frac{2}{3} \frac{d v}{dt} = \frac{2}{5} \frac{k}{L} \frac{\rho_\mu}{\rho_m}\,.
\ee
In both limits one has
\be
v^2 \lsim H L v \sim k \frac{\rho_\sigma + \rho_\mu}{\rho_m}\,,
\label{limit}
\ee
where we have taken into account that $v \lsim LH$ since the characteristic velocity does not change abruptly.

\section{Implications for dark energy}

The amplitude of the fractional energy density fluctuations, $\delta$, associated with domain walls, on a physical scale, $L_V$, much larger than the characteristic scale, $L$, of a domain wall network, is given approximately by
\be
\delta \equiv \frac{\delta \rho_\sigma}{\rho_c} \sim \frac{\Omega_\sigma}{\sqrt N}\,,
\ee
where $N \sim (L_V/L)^3$ is the number of domain walls on a volume $V=L_V^3$, $\Omega_\sigma = \rho_\sigma/\rho_c$, $\rho_c$ is the critical density and $\delta \rho_\sigma$ is the root mean square fluctuation of the domain wall energy density on the physical scale $L_V$. The amplitude of CMB temperature fluctuations generated by domain walls, around the present time, is constrained to be smaller than $10^{-5}$ down to scales of the order of $L_V=H_0^{-1}/100$ (in order not to spoil the agreement between the theoretical model and observations \cite{Komatsu:2008hk,Frieman:2008sn}). This implies 
\be
\delta  \sim 10^3 \, \Omega_{\sigma 0}  \, (H_0 L_0)^{3/2} \lsim 10^{-5}\,.
\ee
Consequently, $H_0 L_0 \lsim 10^{-5}  \, \Omega_{\sigma 0}^{-2/3}$ which results in the constraint $L_0 \lsim 10 \, \Omega_{\sigma 0}^{-2/3} \, {\rm kpc} $. Unless there is an abrupt velocity change, this also translates into a stringent limit on the characteristic velocity of the domain walls at the present time, $v_0 \lsim H_0 L_0 \lsim 10^{-5} \, \Omega_{\sigma 0}^{-2/3}$. Hence, using Eq.  (\ref{limit}) one obtains
\be
k \frac{\rho_{\sigma 0} + \rho_{\mu 0}}{\rho_{m 0}} \lsim (H_0 L_0)^2 \lsim  10^{-10}  \, \Omega_{\sigma 0}^{-4/3}  \,.
\label{limit1}
\ee
The value of the curvature parameter has been estimated using high-resolution numerical simulations of domain wall evolution. For standard domain wall networks without junctions $k \sim 1$ while for domain wall networks with junctions a smaller value has been  observed, but still of order unity \cite{Avelino:2008ve}. A value of $k \ll 1$ is only expected in the case of very special configurations such as 2-dimensional hexagonal lattices with Y-type junctions, square lattices with X-type junctions and triangular lattices with $*$-type junctions. These are very unnatural configurations corresponding to very specific initial conditions which would violate causality if they were to extend over scales larger than the particle horizon. Consequently we shall assume that $k \sim 1$.

Note that domain walls would need to have an average energy density of the order of the critical density to provide a 
significant contribution to the dark energy. We conclude, from Eq. (\ref{limit1}), that the average energy density of the monopole-type junctions would have to be 10 orders of magnitude greater than that. Such a high value of the energy density is in complete disagreement with all cosmological evidence. Of course, if $\Omega_{\sigma 0}$ was very small then frustration could effectively occur but, in that case, domain walls would not play any relevant role as a dark energy component.

\section{Conclusions}

In this paper we have imposed strong constraints on the characteristic length and velocity of domain wall 
networks with string and monopole-type junctions using a semi-analitical VOS model. We have shown that 
a successful domain wall scenario for dark energy would require that $ L_0 \lsim 10 \, {\rm kpc}$ and 
$v_0 \lsim 10^{-5}$. We have demonstrated that such small values of $L_0$ and $v_0$ 
could only be achieved if the contribution of the monopole-type junctions to the total density of the universe 
was several orders of magnitude larger than that of domain walls and strings (assuming 
$\Omega_{\sigma 0} \sim 1$), in complete disagreement with observations.

These results highlight the main difficulty with alternative mechanisms for the frustration of domain wall networks. The inclusion 
of additional degrees of freedom such as heavy junctions and friction may slightly reduce the characteristic length and 
velocity of the domain walls but is insufficient to lead to frustration, due to the limited amount of matter with which 
domain walls can interact while conserving energy and momentum at the present time (the mechanism for the frustration of domain wall networks proposed in \cite{Battye:2009ac} is expected to face similar problems). Our results provide further 
evidence for the absence of a successful mechanism which can lead to the frustration of cosmologically relevant 
domain wall networks.

%%%%%%%%%%%%%%%%%%%%%%%%%%%%%%%%%%%%%%%%%%%%%%%%%%%%%
\begin{acknowledgments}

We thank Joana Oliveira and Roberto Menezes for useful discussions. This work was funded by FCT (Portugal) through contract CERN/FP/109306/2009.

\end{acknowledgments}

%%%%%%%%%%%%%%%%%%%%%%%%%%%%%%%%%%%%%%%%%%%%%%%%%%%%%%%%%%

\bibliography{junctions}

\end{document}